\pdfoutput=1
\documentclass[10pt,a4paper,english,journal]{IEEEtran}
\usepackage[T1]{fontenc}
\usepackage[latin9]{inputenc}
\usepackage{array}
\usepackage{multirow}
\usepackage{amsmath}
\usepackage{amssymb}
\usepackage{stackrel}
\usepackage{graphicx}

\makeatletter

\pdfpageheight\paperheight
\pdfpagewidth\paperwidth

\providecommand{\tabularnewline}{\\}

\usepackage{xcolor,soul}
\sethlcolor{lightgray}

\newcommand{\hilight}[1]{\colorbox{lightgray}{#1}}

\makeatother

\usepackage{babel}
\begin{document}

\title{Inter-Cell Radio Frame Coordination Scheme Based on Sliding Codebook
for 5G TDD Systems}

\author{\IEEEauthorblockN{Ali A. Esswie$^{1,2}$,\textit{ Member, IEEE}, and\textit{ }Klaus
I. Pedersen\textit{$^{1,2}$, Senior Member, IEEE}\\
$^{1}$Nokia Bell-Labs, Aalborg, Denmark\\
$^{2}$Department of Electronic Systems, Aalborg University, Denmark}}

\maketitle
$\pagenumbering{gobble}$
\begin{abstract}
The fifth generation (5G) of the wireless communication networks supports
wide diversity of service classes, leading to a highly dynamic uplink
(UL) and downlink (DL) traffic asymmetry. Thus, dynamic time division
duplexing (TDD) technology has become of a significant importance,
due to its radio frame flexibility. However, fully dynamic TDD systems
suffer from potentially severe inter-cell cross link interference
(CLI). In this paper, we propose a novel inter-cell radio frame coordination
(RFC) scheme based on sliding codebook for fully dynamic TDD 5G networks.
Proposed coordination scheme simultaneously addresses two optimization
objectives of minimizing the average CLI while reliably maximizing
the achievable DL/UL capacity, by virtually extending the RFC degrees
of freedom through a sliding phase-offset RFC codebook design. Compared
to the state-of-the-art TDD studies, the proposed scheme shows significantly
improved ergodic capacity, i.e., at least $\sim140\%$ gain under
both the TCP and UDP protocols, and with much less signaling overhead,
limited to $\textnormal{B-bit}$. The paper offers valuable insights
about how to most efficiently pre-mitigate potential CLI in Macro
TDD systems.

\end{abstract}

\section{Introduction}

\IEEEPARstart{T}{}ime division duplexing (TDD) technology has drawn
a major research attention since day-one of the long term evolution
(LTE) development. The 3rd generation partnership project (3GPP) LTE-Advanced
Rel-12 introduces an enhanced interference mitigation and traffic
adaptation (eIMTA) {[}1, 2{]} to offer a more flexible TDD adaptation.
eIMTA supports seven different TDD radio frame configuration (RFC)
patterns with different downlink (DL) to uplink (UL) traffic ratios,
where each cell dynamically in time adapts its radio frame based on
its own link direction selection criteria, e.g., aggregated traffic
demand {[}3{]}. Though, the fastest possible RFC adaptation periodicity
of eIMTA is the LTE radio frame, i.e., 10 ms.

With the agile frame structure of the 5G new radio (5G-NR) {[}4{]},
e.g., the flexible TDD slot formats and the variable transmission
time interval (TTI) duration, a fully dynamic TDD with much faster
and flexible adaptation becomes feasible. Accordingly, the link direction
switching periodicity can be slot-based, i.e., $\leq1$ ms, instead
of being RFC-based. Thus, 5G-NR TDD systems significantly improve
the spectrum utilization and the ergodic capacity for services with
fast-varying and asymmetric DL and uplink UL traffic {[}5{]}. However,
the coexistence of different link directions over same frequency resources
in adjacent cells results in potential cross link interference (CLI),
i.e., user-to-user (UE-UE), and base-station to base-station (BS-BS)
interference {[}6{]}. In Macro deployments, the CLI, especially the
BS-BS interference, is a critical problem due to the UL and DL power
imbalance. Consequently, the gains of the adaptive RFCs in TDD may
completely vanish due to severe CLI {[}7{]}. 

The state-of-the-art CLI suppression proposals from the open literature
consider either CLI avoidance or post-cancellation. In {[}8{]}, the
combination of cell muting, liquid clustering and enhanced UL power
control is suggested to minimize the average UE-UE and BS-BS CLI.
Additionally, joint UE scheduling and advanced beam-forming techniques
{[}9, 10{]} are envisioned as beneficial to counteract the CLI. Furthermore,
a performance case study on the interaction of the transmission control
protocol (TCP) with the 5G-NR TDD systems is presented {[}11{]}. In
{[}12{]}, a recent proposal introduces perfect CLI cancellation using
full packet exchange, where DL-heavy cells signal neighboring UL-heavy
cells with their respective DL UE transmission information for the
UL-heavy cells to optimally cancel the critical BS-BS CLI.

Compared to the state-of-the-art coordinated TDD studies, significant
inter-cell control signaling overhead and/or ideal periodic UE CLI
measurements are usually assumed, which are infeasible in practice.
Consequently, the overall capacity gains from the TDD RFC flexibility
can be greatly limited due to cell muting or the abrupt changes in
the joint scheduling decisions. Needless to say, an efficient and
flexible coordination scheme is vital for macro TDD systems, to simultaneously
improve the overall ergodic capacity in both UL and DL directions
and with limited signaling overhead. 

In this work, we propose an RFC based sliding codebook (RFCbCB) coordination
scheme for 5G TDD systems. The proposed scheme effectively boosts
the TDD system degrees of freedom, coming from its frame flexibility,
with the size of a specially pre-designed RFC codebook. Consequently,
the maximum possible ergodic capacity is achieved while simultaneously
guaranteeing acceptable CLI levels and with a significantly reduced
inter-cell control signaling overhead, limited to $\textnormal{B-bit}$.
Extensive system level simulations show that the proposed RFCbCB scheme
significantly improves the ergodic capacity by the efficient CLI avoidance
in both DL and UL directions simultaneously. Moreover, as various
applications require different link reliability levels, e.g., TCP
is commonly used with the 5G-NR enhanced mobile broadband service
class and user data-gram protocol (UDP) with latency critical traffic,
we evaluate the proposed scheme performance over both transport protocols
to study the effect of the CLI on the TCP flow and congestion controls,
respectively.

Due to the complexity of the 5G-NR {[}4{]} and addressed problem herein,
we evaluate the performance by extensive system simulations, where
the main TDD functionalities are calibrated against the 3GPP 5G-NR
assumptions. This includes UL and DL channel modeling, dynamic modulation
and coding schemes (MCS), dynamic hybrid automatic repeat request
(HARQ), and dynamic UE scheduling. 

This paper is organized as follows. Section II presents the system
model of this work. Section III introduces the problem formulation
while Section IV details the proposed RFCbCB coordination scheme and
the reference studies to compare against. Section V discusses the
performance evaluation results and paper is finally concluded in Section
VI.

\section{System Model }

We consider a 5G-NR TDD system with a single cluster of $C$ cells,
each with $N_{t}$ antennas. Each cell serves an average of $K^{\textnormal{dl}}$
and $K^{\textnormal{ul}}$ uniformly-distributed DL and UL UEs, each
with $M_{r}$ antennas. Without loss of generality, we assume the
FTP3 traffic modeling with finite payload sizes $\textnormal{\ensuremath{\mathit{f}^{dl}}}$
and $\textnormal{\ensuremath{\mathit{f}^{ul}}}$ bits, and Poisson
point arrival processes $\textnormal{\ensuremath{\lambda}}^{\textnormal{dl}}$
and $\textnormal{\ensuremath{\lambda}}^{\textnormal{ul}},$ in the
DL and UL directions, respectively. Thus, the total offered traffic
load per cell in DL direction is: $K^{\textnormal{dl}}\times\textnormal{\ensuremath{\mathit{f}^{dl}}}\times\textnormal{\ensuremath{\lambda}}^{\textnormal{dl}}$
and in UL direction: $K^{\textnormal{ul}}\times\textnormal{\ensuremath{\mathit{f}^{ul}}}\times\textnormal{\ensuremath{\lambda}}^{\textnormal{ul}}$,
respectively. 

In the time domain, we assume an RFC of 10 sub-frames, each is 1-ms
and can be either a DL, UL or special sub-frame. In the frequency
domain, UEs are dynamically multiplexed by the orthogonal frequency
division multiple access (OFDMA), with the smallest schedulable unit
as the physical resource block (PRB) of 12-subcarriers, each is 15
kHz. Thus, a sub-frame is one slot of 14-OFDM symbols. Nonetheless,
the proposed solution is also valid with different numerologies of
the 5G-NR sub-carrier spacing, TTI duration, and number of TDD slots
per sub-frame, respectively. 

Within each cluster, an arbitrary master cell is declared where other
cells act as slaves. Such master cell can be manually pre-configured
since it is independent from time and the coordination technology.
All cells within each cluster are assumed to be bi-bidirectionally
inter-connected to the master cell through the \textit{Xn interface},
as shown in Fig. 1. 

\begin{figure}
\begin{centering}
\includegraphics[scale=0.7]{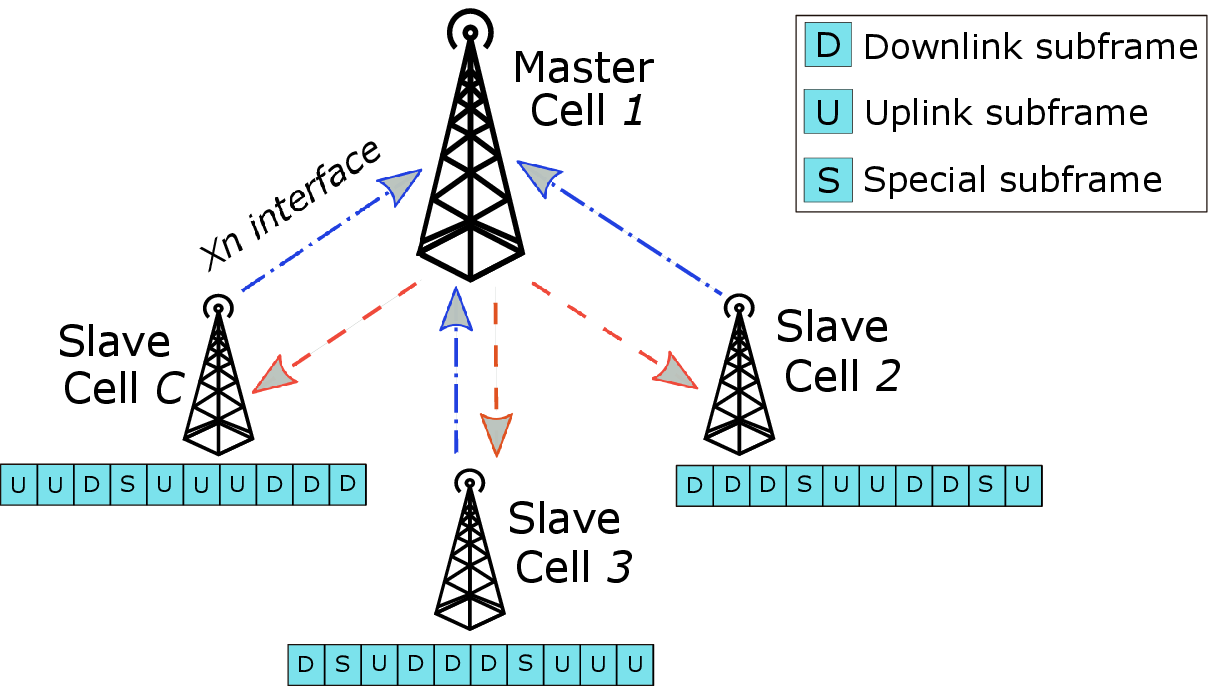}
\par\end{centering}
\centering{}{\small{}Fig. 1. Proposed RFCbCB: system model.}{\small \par}
\end{figure}

Let $\text{\ensuremath{\mathfrak{B}}}_{\textnormal{dl}},$ $\text{\ensuremath{\mathfrak{B}}}_{\textnormal{ul}}$,
$\text{\ensuremath{\mathcal{K}}}_{\textnormal{dl}}$ and $\text{\ensuremath{\mathcal{K}}}_{\textnormal{ul}}$
indicate the sets of cells and UEs in the DL and UL transmission modes,
respectively. Then, the received signal at the $k^{th}$ UE, where
$k\text{\ensuremath{\in\text{\ensuremath{\mathcal{K}}}_{\textnormal{dl}}}}$,
$c_{k}\text{\ensuremath{\in\text{\ensuremath{\mathfrak{B}}}_{\textnormal{dl}}}}$,
is 

\begin{equation}
\boldsymbol{\textnormal{y}}_{k,c_{k}}^{\textnormal{dl}}=\underbrace{\boldsymbol{\textnormal{\textbf{H}}}_{k,c_{k}}^{\textnormal{dl}}\boldsymbol{\textnormal{\textbf{v}}}_{k}s_{k}}_{\text{Useful signal}}+\underbrace{\sum_{i\in\text{\ensuremath{\mathcal{K}}}_{\textnormal{dl}}\backslash k}\boldsymbol{\textnormal{\textbf{H}}}_{k,c_{i}}^{\textnormal{dl}}\boldsymbol{\textnormal{\textbf{v}}}_{i}s_{i}}_{\text{BS to UE interference}}+\underbrace{\sum_{j\in\text{\ensuremath{\mathcal{K}}}_{\textnormal{ul}}}\boldsymbol{\textnormal{\textbf{G}}}_{k,j}\boldsymbol{\textnormal{\textbf{w}}}_{j}s_{j}}_{\text{UE to UE interference}}+\boldsymbol{\textnormal{\textbf{n}}}_{k}^{\textnormal{dl}},
\end{equation}
where $\boldsymbol{\textnormal{\textbf{H}}}_{k,c_{i}}^{\textnormal{dl}}\in\text{\ensuremath{\mathcal{C}}}^{M_{r}\times N_{t}}$
denotes the DL channel from the cell, serving the $i^{th}$ UE, to
the $k^{th}$ UE, $\boldsymbol{\textnormal{\textbf{v}}}_{k}\in\text{\ensuremath{\mathcal{C}}}^{N_{t}\times1}$
and $s_{k}$ are the single-stream precoding vector at the $c_{k}^{th}$
cell and data symbol of the $k^{th}$ UE, respectively, $\boldsymbol{\textnormal{\textbf{G}}}_{k,j}\in\text{\ensuremath{\mathcal{C}}}^{M_{r}\times M_{r}}$
is the channel between the $k^{th}$ and $j^{th}$ UEs. $\boldsymbol{\textnormal{\textbf{w}}}_{j}\in\text{\ensuremath{\mathcal{C}}}^{M_{r}\times1}$
is the single-stream precoding vector at the $j^{th}$ UE, and $\boldsymbol{\textnormal{\textbf{n}}}_{k}^{\textnormal{dl}}$
is the additive white Gaussian noise at the $k^{th}$ UE. The first
summation denotes the DL-to-DL inter-cell interference while the second
summation represents the inter-cell UE-UE CLI. Similarly, the received
signal at the $c_{k}^{th}$ cell, where $c_{k}\text{\ensuremath{\in\text{\ensuremath{\mathfrak{B}}}_{\textnormal{ul}}}}$
from $k\text{\ensuremath{\in\text{\ensuremath{\mathcal{K}}}_{\textnormal{ul}}}},$
is 

\begin{equation}
\boldsymbol{\textnormal{y}}_{c_{k},k}^{\textnormal{ul}}=\underbrace{\boldsymbol{\textnormal{\textbf{H}}}_{c_{k},k}^{\textnormal{ul}}\boldsymbol{\textnormal{\textbf{w}}}_{k}s_{k}}_{\text{Useful signal}}+\underbrace{\sum_{j\in\text{\ensuremath{\mathcal{K}}}_{\textnormal{ul}}\backslash k}\boldsymbol{\textnormal{\textbf{H}}}_{c_{k},j}^{\textnormal{dl}}\boldsymbol{\textnormal{\textbf{w}}}_{j}s_{j}}_{\text{UE to BS interference}}+\underbrace{\sum_{i\in\text{\ensuremath{\mathcal{K}}}_{\textnormal{dl}}}\boldsymbol{\textnormal{\textbf{Q}}}_{c_{k},c_{i}}\boldsymbol{\textnormal{\textbf{v}}}_{i}s_{i}}_{\text{BS to BS interference}}+\boldsymbol{\textnormal{\textbf{n}}}_{c_{k}}^{\textnormal{ul}},
\end{equation}
where $\boldsymbol{\textnormal{\textbf{Q}}}_{c_{k},c_{i}}\in\text{\ensuremath{\mathcal{C}}}^{N_{t}\times N_{t}}$
is the channel between the cells that serve the $k^{th}$ and $i^{th}$
UEs, respectively, where $k\text{\ensuremath{\in\text{\ensuremath{\mathcal{K}}}_{\textnormal{ul}}}}$
and $i\in\text{\ensuremath{\mathcal{K}}}_{\textnormal{dl}}$. The
first summation implies the UL-to-UL inter-cell interference while
the second summation denotes the inter-cell BS-BS CLI. Then, the received
signal is decoded using the linear minimum mean square error interference
rejection combining receiver (LMMSE-IRC) {[}4{]} matrix $\boldsymbol{\textnormal{\textbf{a}}}$
as

\begin{equation}
\hat{s}_{k}^{\kappa}=\left(\boldsymbol{\textnormal{\textbf{a}}}_{k}^{\kappa}\right)^{\textnormal{H}}\boldsymbol{\textnormal{y}}_{k}^{\kappa},
\end{equation}
where $\text{\ensuremath{\mathcal{X}}}^{\kappa},\kappa\text{\ensuremath{\in}}\{\textnormal{ul},\textnormal{dl}\}$,
and $\left(\bullet\right)^{\textnormal{H}}$ denotes the Hermitian
operation. Finally, the received signal-to-interference-noise-ratio
(SINR) levels in the DL direction at the $k^{th}$ UE and in the UL
direction at the $c_{k}^{th}$ cell, respectively, are expressed by

{\small{}
\begin{equation}
\gamma_{k}^{\textnormal{dl}}=\frac{p_{c_{k}}^{\textnormal{dl}}\left\Vert \boldsymbol{\textnormal{\textbf{H}}}_{k,c_{k}}^{\textnormal{dl}}\boldsymbol{\textnormal{\textbf{v}}}_{k}\right\Vert ^{2}}{\sigma^{2}+\underset{i\in\text{\ensuremath{\mathcal{K}}}_{\textnormal{dl}}\backslash k}{\sum}p_{c_{i}}^{\textnormal{dl}}\left\Vert \boldsymbol{\textnormal{\textbf{H}}}_{k,c_{i}}^{\textnormal{dl}}\boldsymbol{\textnormal{\textbf{v}}}_{i}\right\Vert ^{2}+\underset{j\in\text{\ensuremath{\mathcal{K}}}_{\textnormal{ul}}}{\sum}p_{j}^{\textnormal{ul}}\left\Vert \boldsymbol{\textnormal{\textbf{G}}}_{k,j}\boldsymbol{\textnormal{\textbf{w}}}_{j}\right\Vert ^{2}},
\end{equation}
}{\small \par}

{\small{}
\begin{equation}
\gamma_{c_{k}}^{\textnormal{ul}}=\frac{p_{k}^{\textnormal{ul}}\left\Vert \boldsymbol{\textnormal{\textbf{H}}}_{c_{k},k}^{\textnormal{ul}}\boldsymbol{\textnormal{\textbf{w}}}_{k}\right\Vert ^{2}}{\sigma^{2}+\underset{j\in\text{\ensuremath{\mathcal{K}}}_{\textnormal{ul}}\backslash k}{\sum}p_{j}^{\textnormal{ul}}\left\Vert \boldsymbol{\textnormal{\textbf{H}}}_{c_{k},j}^{\textnormal{dl}}\boldsymbol{\textnormal{\textbf{w}}}_{j}\right\Vert ^{2}+\underset{i\in\text{\ensuremath{\mathcal{K}}}_{\textnormal{dl}}}{\sum}p_{c_{i}}^{\textnormal{dl}}\left\Vert \boldsymbol{\textnormal{\textbf{Q}}}_{c_{k},c_{i}}\boldsymbol{\textnormal{\textbf{v}}}_{i}\right\Vert ^{2}},
\end{equation}
}where $p_{c_{k}}^{\textnormal{dl}}$ and $p_{k}^{\textnormal{ul}}$
are the transmission power of the $c_{k}^{th}$ cell in the DL direction
and the $k^{th}$ UE in the UL direction, respectively. As can be
observed from (5), the BS-BS CLI can significantly degrade the perceived
UL SINR level due to the DL and UL power imbalance, i.e., $p_{c_{i}}^{\textnormal{dl}}\gg p_{k}^{\textnormal{ul}}.$ 

\section{Problem Formulation - CLI Mitigation }

In fully TDD systems, cells may not adopt exactly the same RFC. Thus,
neighboring cells experience different transmission directions over
several sub-frames, causing severe BS-BS and UE-UE CLI. Accordingly,
the lower-power UL transmissions are severely degraded due to the
strong CLI resulting from adjacent larger-power DL transmissions.
As a result, the achievable UL capacity exhibits a significant loss,
leading to more buffered UL traffic in those victim cells. Hence,
these cells will be dictated by new and buffered UL traffic leading
to a limited DL capacity and a highly degraded overall spectral efficiency
as a consequence. 

To address this problem, the proposed RFCbCB seeks to maximize the
long-term ergodic capacity while simultaneously preserving limited
inter-cell sub-frame misalignment, thus, an acceptable average CLI.
Let $u_{c}$ and $d_{c}$ denote the estimated numbers of UL and DL
sub-frames in an arbitrary RFC while $u_{c}^{\textnormal{opt.}}$
and $d_{c}^{\textnormal{opt.}}$ indicate the corresponding optimal
numbers. Thus, we define a heuristic optimization problem as 
\begin{figure}
\begin{centering}
\includegraphics[scale=0.56]{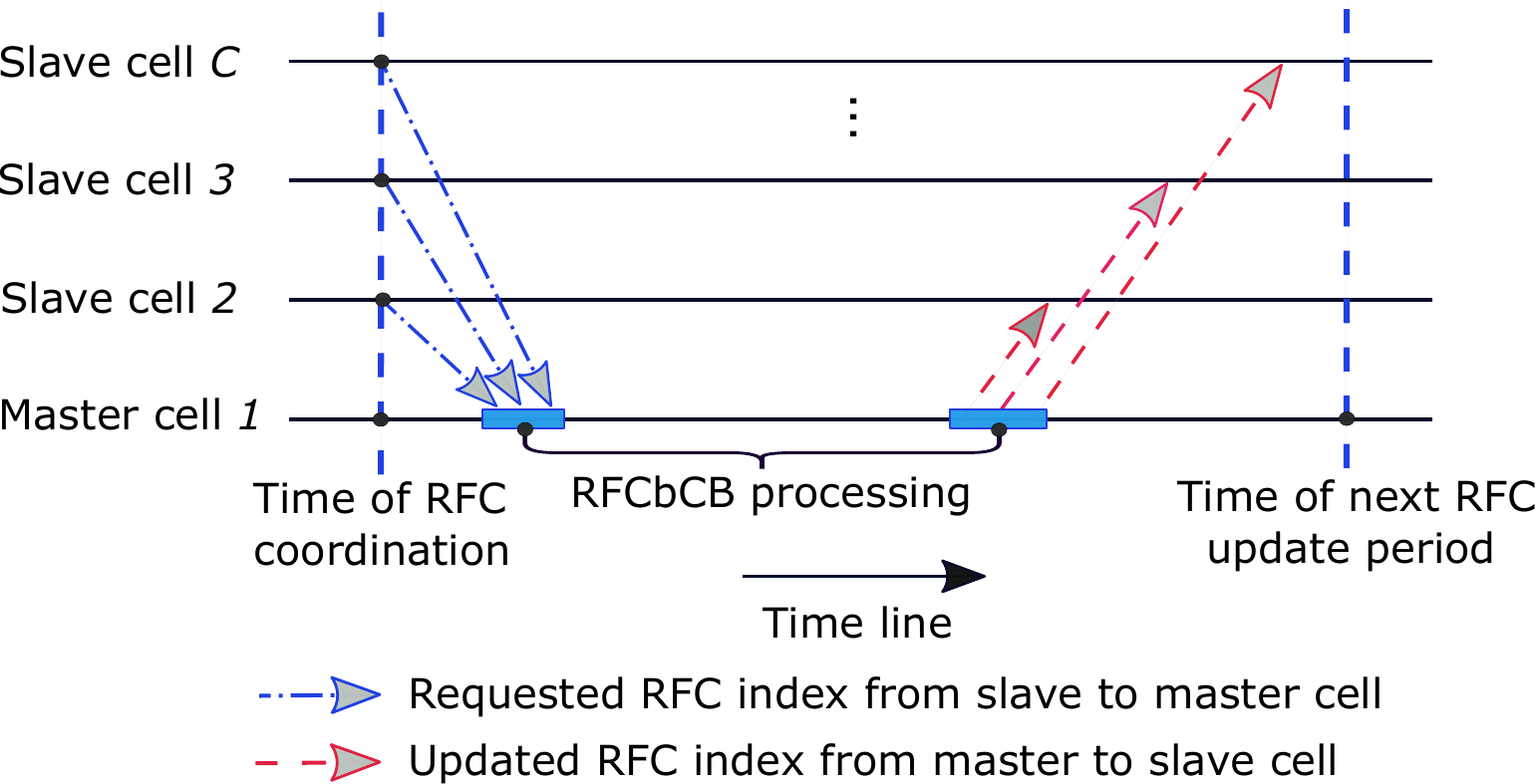}
\par\end{centering}
\centering{}{\small{}Fig. 2. Timing diagram of the inter-cell RFC
coordination. }{\small \par}
\end{figure}

\begin{equation}
R\triangleq\underset{c}{\arg\max}\stackrel[c=1]{C}{\sum}\min\left(\textnormal{\ensuremath{u_{c}},\ensuremath{u_{c}^{\textnormal{opt.}}}}\right)\digamma_{c}^{u}+\min\left(\textnormal{\ensuremath{d_{c}},\ensuremath{d_{c}^{\textnormal{opt.}}}}\right)\digamma_{c}^{d},
\end{equation}

\begin{equation}
\phi_{c}\left(\eta_{c}\right)=\underset{x}{\arg\min}\,\frac{1}{C}\stackrel[x=1,x\neq c]{C}{\sum}\varphi_{c,x}\left(\eta_{c},\eta_{x}\right),
\end{equation}
where $R$ is the aggregate capacity of the cluster, $\digamma_{c}^{u}$
and $\digamma_{c}^{d}$ are the rate utility functions of the UL and
DL transmissions in the $c^{th}$ cell, i.e., the achievable capacity
gain from having either UL or DL transmission. $\phi_{c}\left(\eta_{c}\right)$
and $\varphi_{c,x}\left(\eta_{c},\eta_{x}\right)$ are the average
and actual sub-frame misalignment of the RFC requested by the $c^{th}$
cell $\eta_{c}$ and between the RFCs of the $c^{th}$ and $x^{th}$
cells, i.e., $\eta_{c}$ and $\eta_{x}$, respectively, $\forall x\neq c$.
To maximize (6), $u_{c}=u_{c}^{\textnormal{opt.}}$ and $d_{c}=d_{c}^{\textnormal{opt.}}$
should be preserved; however, $u_{c}^{\textnormal{opt.}}$ and $d_{c}^{\textnormal{opt.}}$
may result in a large sub-frame misalignment, leading to severe CLI
in the cluster and the overall capacity $R$ shall be significantly
degraded accordingly. 

\section{Proposed RFCbCB Coordination }

A specially designed RFC CB is pre-defined and assumed pre-known to
all cells in each cluster. Slightly before each RFC update periodicity,
each slave cell identifies its desired upcoming RFC from the CB that
satisfies its link direction selection objectives. Next, it signals
the master cell with the index of its desired RFC from the CB over
\textit{Xn interface}. The master BS then seeks to simultaneously
satisfy both (6) and (7). Hence, the master cell may slightly change
the RFC indices, which were requested by slave cells. Finally, it
signals the updated RFC indices back to the slave cells, which should
be used during the next RFC update period. Fig. 2 depicts the generic
timing diagram of the proposed solution. 

\subsection{Proposed Inter-Cell Coordination Scheme}

\textbf{\textit{The design of the RFC sliding codebook:}}

A pre-defined RFC CB of size $\mathcal{N}$ unique RFCs is constructed,
where it is divided into $L$ different sub-CBs. Each sub-CB contains
RFCs with the same DL:UL sub-frame ratio in a radio frame, i.e., $d_{c}:u_{c};$
however, with a different DL and UL sub-frame placement as depicted
in Fig. 3, where each RFC is cyclic-shift of the other RFCs in the
same sub-CB. The total number of RFCs, sub-CBs, and cyclic-shift in
the CB are arbitrary design parameters. 

\begin{figure}
\begin{centering}
\includegraphics[scale=0.4]{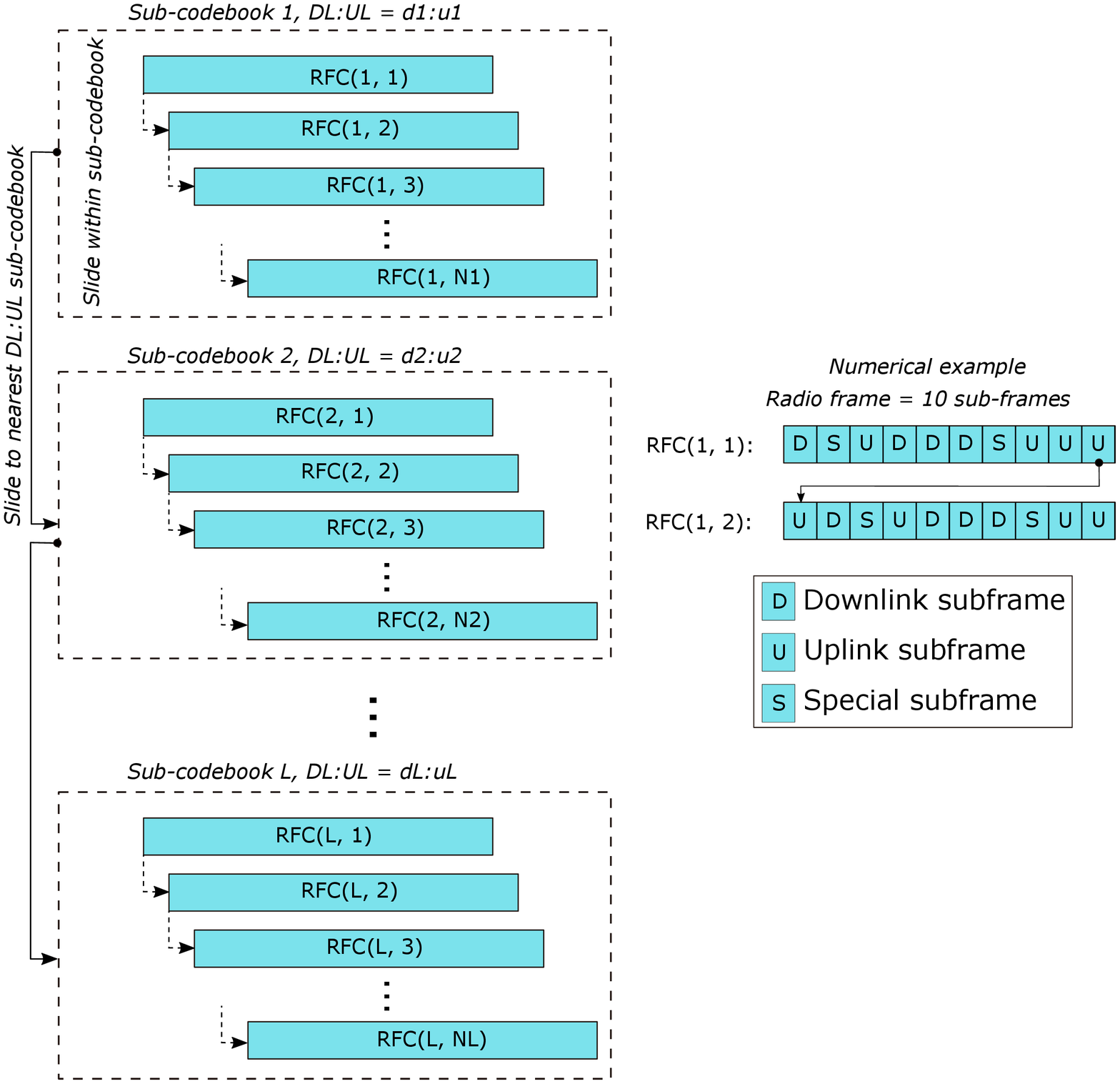}
\par\end{centering}
\centering{}{\small{}Fig. 3. Proposed RFCbCB: RFC sliding CB design.}{\small \par}
\end{figure}

\textbf{\textit{At the slave cells within a cluster:}}

At each RFC update period, each slave cell selects an RFC from the
CB that most meets its own link direction selection criteria. Without
loss of generality, we consider the buffered traffic including pending
re-transmissions as the main criterion with which each cell determines
its required $d_{c}:u_{c}$ ratio. Let $\beta_{c}$ implies the traffic
load threshold of the $c_{k}^{th}$ cell, as

\begin{equation}
\beta_{c}\leq\frac{\sum Z_{c}^{\textnormal{dl}}}{\sum Z_{c}^{\textnormal{dl}}+\sum Z_{c}^{\textnormal{ul}}},
\end{equation}
where $\sum Z_{c}^{\textnormal{dl}}$ and $\sum Z_{c}^{\textnormal{ul}}$
are the total buffered traffic in the DL and UL directions, respectively.
The traffic load threshold $\beta_{c}$ is used to bias the link direction
selection to either DL or UL. For an instance, with fair $\beta_{c}=0.5$,
if $\sum Z_{c}^{\textnormal{dl}}\geq\sum Z_{c}^{\textnormal{ul}}$,
a cell decides a DL-heavy RFC. Finally, slave cells feedback the master
cell with $\textnormal{B}=\log_{2}\left(\text{\ensuremath{\mathcal{N}}}\right)$
bits on the \textit{Xn interface} to indicate the index of the desired
RFC from the CB. 

\textbf{\textit{At the master cell within a cluster:}}

The master cell first identifies the \textit{common RFC}, as the most
requested RFC by the majority of the slave cells in the cluster. If
not possible, the master cell considers any requested RFC as the common
RFC of this cluster, and to which other requested RFCs from other
slave cells must satisfy the minimum possible sub-frame misalignment
with. 

Then, for each requested RFC $\eta_{c}$ of the $c_{k}^{th}$ cell,
the master cell calculates the sub-frame misalignment to the common
RFC $\delta_{x},\forall x\neq c$ as: $\varphi_{c,x}\left(\eta_{c}\left(u_{c},d_{c}\right),\delta_{x}\left(u_{x},d_{x}\right)\right)$.
If $\varphi_{c,x}\leq\psi,$ with $\psi$ as a pre-defined sub-frame
misalignment threshold, the master cell skips updating such requested
RFC, i.e., does not change it, since it originally drives a limited
CLI. Otherwise, the master cell slides over all RFCs within the same
sub-CB of the requested RFC $\eta_{c}$. It calculates the corresponding
misalignment values $\varphi_{c,x}$ and selects the RFC with the
minimum sub-frame misalignment. Finally, it performs two variations:
\begin{itemize}
\item \textbf{Option-1}: if the sub-frame misalignment of the selected RFC
is below $\psi$, the master cell considers such RFC as the updated
RFC of this cell. Hence, an acceptable average CLI level is guaranteed
across the upcoming RFC update period while still preserving the same
required traffic service ratio $d_{c}:u_{c}$, improving both the
slave cells and overall cluster capacity, respectively. 
\item \textbf{Option-2}: if $\varphi_{c,x}\leq\psi$ is not feasible with
all RFCs in the same sub-CB, the master cell slides to a different
sub-CB in the CB, with the nearest $d_{c}:u_{c}$ ratio to the requested
ratio, e.g., $d_{c}:u_{c}=2:6\underset{\textnormal{slide\,to}}{\rightarrow}d_{c}^{'}:u_{c}^{'}=3:5$,
and repeats the same operation. Herein, the master cell slightly sacrifices
part of the full TDD RFC flexibility, due to the change in the requested
$d_{c}:u_{c}.$ However, such capacity loss is bounded over only a
limited number of sub-frames and is reversibly proportional to the
size of the CB, since the master cell slides only to the nearest $d_{c}:u_{c}$
sub-CB. As will be discussed in Section V, this capacity loss is fully
recovered on the long-term statistics due to the significantly reduced
CLI. 
\end{itemize}
As a last resort, if the sub-frame misalignment threshold can not
be further satisfied, the master cell considers the RFC with the minimum
misalignment $\varphi_{c,x}$, from either the same or different sub-CB
as the requested one, as the updated RFC of this slave cell even it
does not satisfy $\varphi_{c,x}\leq\psi$. Finally, the master cell
feeds-back all slave cells back with $\textnormal{B}$-bit indices
over the \textit{Xn interface} to indicate their respective updated
RFCs to be used over the next RFC update period. 

\subsection{Comparison to the state-of-the-art TDD studies}

We compare the performance of the proposed solution against the following
state-of-the-art TDD proposals as:

\textbf{Fully-uncoordinated TDD (FUC)}: all cells in the cluster independently
select their respective RFCs from the CB based on the traffic criterion
in (8). No inter-cell RFC coordination is assumed. Thus, a large inter-cell
sub-frame misalignment and hence, severe CLI levels can be exhibited. 

\textbf{Ideal-UL interference coordination TDD (IUIC)} {[}12{]}: within
a cluster, cells independently select respective RFCs based on (8).
Then, the DL-heavy cells feedback their respective DL payload, PRB
mapping, UE MCS and precoding information to UL-heavy cells over the
\textit{Xn interface}. Accordingly, the UL-heavy cells are perfectly
able to fully suppress the BS-BS CLI, i.e., $\textnormal{BS-BS CLI = 0}$.
Therefore, the IUIC is an UL-optimal TDD coordination scheme; however,
with a significant signaling overhead over the back-haul links. 

\section{Performance Evaluation}

The performance assessment of the proposed coordination scheme  is
based on highly dynamic system level simulations, where the main 3GPP
assumptions are followed {[}4{]}. The major simulation setup parameters
are listed in Table I. At each TTI, each cell dynamically and independently
schedules UEs over system PRBs according to the proportional fair
criterion. Herein, we assume fully dynamic  link adaptation and Chase
combining HARQ, respectively, where the DL/UL HARQ feedback is sent
with a higher priority during the first available transmission opportunity
of the adopted RFC. The sub-carrier SINR level is calculated using
the LMMSE-IRC receiver. For MCS selection, sub-carrier SINR levels
are combined using the effective exponential SNR mapping algorithm
to obtain an effective wide-band SINR. Finally, we evaluate the performance
of the proposed RFCbCB scheme under both TCP and UDP, with different
offered traffic loads per cell and for the two proposed options. 

\begin{table}
\caption{{\small{}Simulation parameters.}}
\centering{}%
\begin{tabular}{c|c}
\hline 
Parameter & Value\tabularnewline
\hline 
Environment & 3GPP-UMA, one cluster, 21 cells\tabularnewline
\hline 
UL/DL channel bandwidth & 10 MHz, TDD\tabularnewline
\hline 
TDD mode & Synchronized \tabularnewline
\hline 
Antenna setup & $N_{t}=8$ Tx, $M_{r}=2$ Rx\tabularnewline
\hline 
Average user load & $K^{\textnormal{dl}}=K^{\textnormal{ul}}=$ 10 users per cell\tabularnewline
\hline 
UL/DL receiver & LMMSE-IRC\tabularnewline
\hline 
Traffic model & $\begin{array}{c}
\textnormal{FTP3, \textnormal{\ensuremath{\mathit{f}^{dl}}} = \textnormal{\ensuremath{\mathit{f}^{ul}}} = 4000 bits}\\
\textnormal{\ensuremath{\textnormal{\ensuremath{\lambda}}^{\textnormal{dl}}} = 500, 375, and 250 pkts/sec}\\
\textnormal{\ensuremath{\textnormal{\ensuremath{\lambda}}^{\textnormal{ul}}} = 500, 375, and 250 pkts/sec}
\end{array}$\tabularnewline
\hline 
User scheduler & Proportional fair\tabularnewline
\hline 
$\begin{array}{c}
\textnormal{\textnormal{Offered average load per cell}}\\
\textnormal{DL:UL}
\end{array}$ & $\begin{array}{c}
\textnormal{\textnormal{DL:UL = 2:1 (20:10)} Mbps}\\
\textnormal{\textnormal{DL:UL = 1:1 (15:15)} Mbps}\\
\textnormal{DL:UL = 1:2 (10:20) Mbps}
\end{array}$\tabularnewline
\hline 
Proposed RFCbCB setup & $\begin{array}{c}
\textnormal{\textnormal{\ensuremath{\psi}\ = 3}}\textnormal{ sub-frames}\\
\textnormal{\ensuremath{\mathcal{N}} = 55}\textnormal{ RFCs}\\
L\textnormal{ = 7}\textnormal{ sub-CBs}\\
\textnormal{\textnormal{ B = 6 bits}}
\end{array}$\tabularnewline
\hline 
Transport layer setup & $\begin{array}{c}
\textnormal{TCP/UDP max PDU: 1500 Bytes}\\
\textnormal{ Congestion control: CUBIC}\\
\textnormal{Slow start threshold: 35 MSS}
\end{array}$\tabularnewline
\hline 
\end{tabular}
\end{table}
 
\begin{table*}
\caption{Proposed RFCbCB (option-1): achievable DL and UL throughput (Mbps)
per cluster, with TCP.}
\centering{}%
\begin{tabular}{c|c|c|c|c|c|c|c}
\hline 
\multirow{2}{*}{Traffic Ratio} & \multirow{2}{*}{Offered load per cell (Mbps)} & \multicolumn{2}{c|}{FUC} & \multicolumn{2}{c|}{Proposed RFCbCB (option-1)} & \multicolumn{2}{c}{IUIC}\tabularnewline
\cline{3-8} 
 &  & DL  & UL & DL  & UL & DL  & UL\tabularnewline
\hline 
DL:UL = 2:1 & DL:UL = 20:10 & $\begin{array}{c}
131.52\\
\hilight{0.0\%}
\end{array}$ & $\begin{array}{c}
17.63\\
\hilight{0.0\%}
\end{array}$ & $\begin{array}{c}
167.87\\
\hilight{+24.2\%}
\end{array}$ & $\begin{array}{c}
109.23\\
\hilight{+144.41\%}
\end{array}$ & $\begin{array}{c}
228.38\\
\hilight{+53.82\%}
\end{array}$ & $\begin{array}{c}
158.61\\
\hilight{+159.98\%}
\end{array}$\tabularnewline
\hline 
DL:UL = 1:1 & DL:UL = 15:15 & $\begin{array}{c}
117.73\\
\hilight{0.0\%}
\end{array}$ & $\begin{array}{c}
26.12\\
\hilight{0.0\%}
\end{array}$ & $\begin{array}{c}
149.53\\
\hilight{+23.79\%}
\end{array}$ & $\begin{array}{c}
154.39\\
\hilight{+142.12\%}
\end{array}$ & $\begin{array}{c}
191.09\\
\hilight{+47.50\%}
\end{array}$ & $\begin{array}{c}
195.16\\
\hilight{+152.78\%}
\end{array}$\tabularnewline
\hline 
DL:UL = 1:2 & DL:UL = 10:20 & $\begin{array}{c}
106.07\\
\hilight{0.0\%}
\end{array}$ & $\begin{array}{c}
138.24\\
\hilight{0.0\%}
\end{array}$ & $\begin{array}{c}
113.10\\
\hilight{+6.41\%}
\end{array}$ & $\begin{array}{c}
198.65\\
\hilight{+35.86\%}
\end{array}$ & $\begin{array}{c}
139.79\\
\hilight{+27.43\%}
\end{array}$ & $\begin{array}{c}
264.6\\
\hilight{+62.73\%}
\end{array}$\tabularnewline
\hline 
\end{tabular}
\end{table*}

Table II shows the achievable DL and UL throughput per cluster under
TCP of the FUC, proposed RFCbCB (option-1), and IUIC, for different
DL:UL traffic ratios. As can be clearly observed, with all traffic
load variations, the proposed RFCbCB (option-1) provides a significant
capacity improvement in both the DL and UL directions, compared to
the FUC. It also approaches the optimal IUIC, due to the significantly
reduced average CLI. For instance, with a BS-BS CLI extreme case,
i.e., DL:UL = 2:1, proposed RFCbCB (option-1) achieves $\sim+144.41\%$
gain in the UL capacity than the FUC. The optimal IUIC offers the
best DL and UL throughput since the BS-BS CLI is assumed perfectly
suppressed. Thus, UL traffic gets transmitted faster with zero CLI,
i.e., UL PRBs become of higher capacity, leaving more time and resources
for DL traffic. Accordingly, both UL and DL capacity are improved.
The FUC exploits the full TDD RFC flexibility; however, the aggregated
capacity is severely degraded due to the exhibited strong CLI levels. 

The RFCbCB performance gain is mainly due to the significant reduction
of the average CLI. Hence, Figs. 4.a and 4.b show the empirical cumulative
distribution function (ECDF) of the BS-BS and UE-UE CLI, respectively,
averaged over all system PRBs with DL:UL = 2:1. The proposed RFCbCB
offers a highly improved CLI performance, i.e., more than 70\% and
50\% of the simulation time are UE-UE and BS-BS CLI-free, respectively.
Compared to the optimal IUIC, the RFCbCB shows a further reduced UE-UE
CLI since IUIC is only UL-optimal with no BS-BS CLI (no ECDF of the
BS-BS CLI with IUIC in Fig. 4.b). However, the proposed RFCbCB non-biasedly
seeks for minimizing both BS-BS and UE-UE CLI, respectively. 

\begin{figure}
\begin{centering}
\includegraphics[scale=0.45]{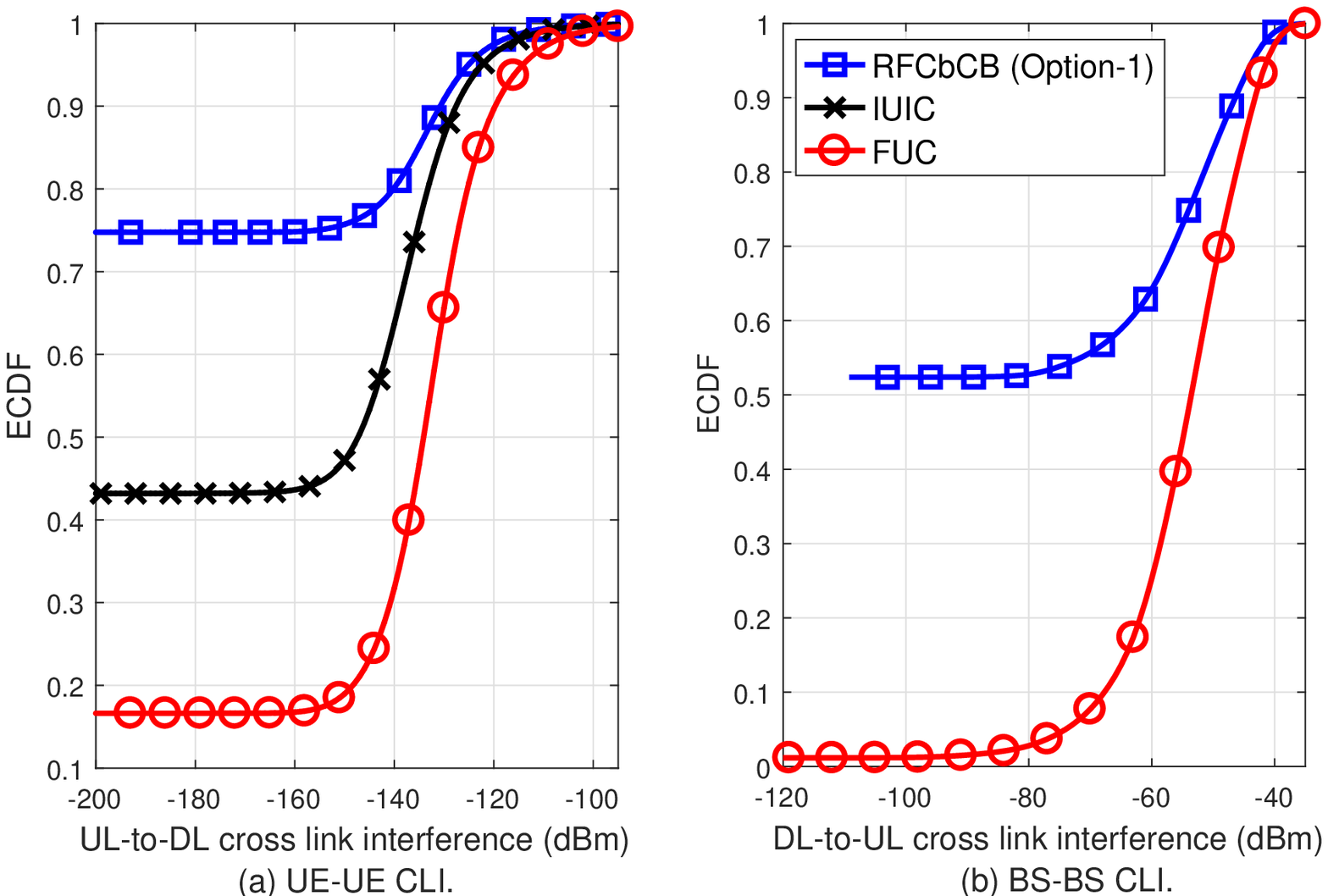}
\par\end{centering}
\centering{}{\small{}Fig. 4. Cross link interference performance (dBm),
with TCP.}{\small \par}
\end{figure}

Looking at the TCP performance, Fig. 5 depicts the ECDF of the average
UL and DL TCP congestion window (CWND) in MB. The TCP CWND is a congestion
control measure, applied at the transmitter side, to counteract network
congestion. It defines the maximum rate bound that a transmitter can
use towards a receiver such that it is exponentially increased when
a successful TCP acknowledgment (ACK) is received, otherwise, it is
decreased. Hence, the TCP transmission rate is restricted by either
the transmitter CWND or the receiver advertised maximum window. Accordingly,
the TCP CWND performance is highly correlated to the exhibited CLI
in TDD systems. As shown in Fig. 5, the FUC inflicts an extremely
small CWND size, due to the exhibited severe CLI. The proposed RFCbCB
achieves $\sim+168.25
$ gain in the 90 percentile CWND size compared to the FUC. However,
the optimal IUIC offers the best average CWND performance, despite
that it exhibits a larger average UE-UE CLI than the proposed RFCbCB,
as shown in Fig. 4.a. This consolidates the fact that the BS-BS CLI
has a stronger impact on overall capacity than the UE-UE CLI due to
the power imbalance between UL and DL transmissions. Though, the proposed
RFCbCB achieves an average $\sim+58.1\%$ gain in the CWND size than
IUIC for the percentiles below 40\%, due to the achievable $\sim-52.1\%$
reduction in the UE-UE CLI as in Fig. 4.a. Thus, with the proposed
RFCbCB, cell-edge DL UEs inflict much less CLI from adjacent inter-cell-edge
UL UEs.

\begin{figure}
\begin{centering}
\includegraphics[scale=0.5]{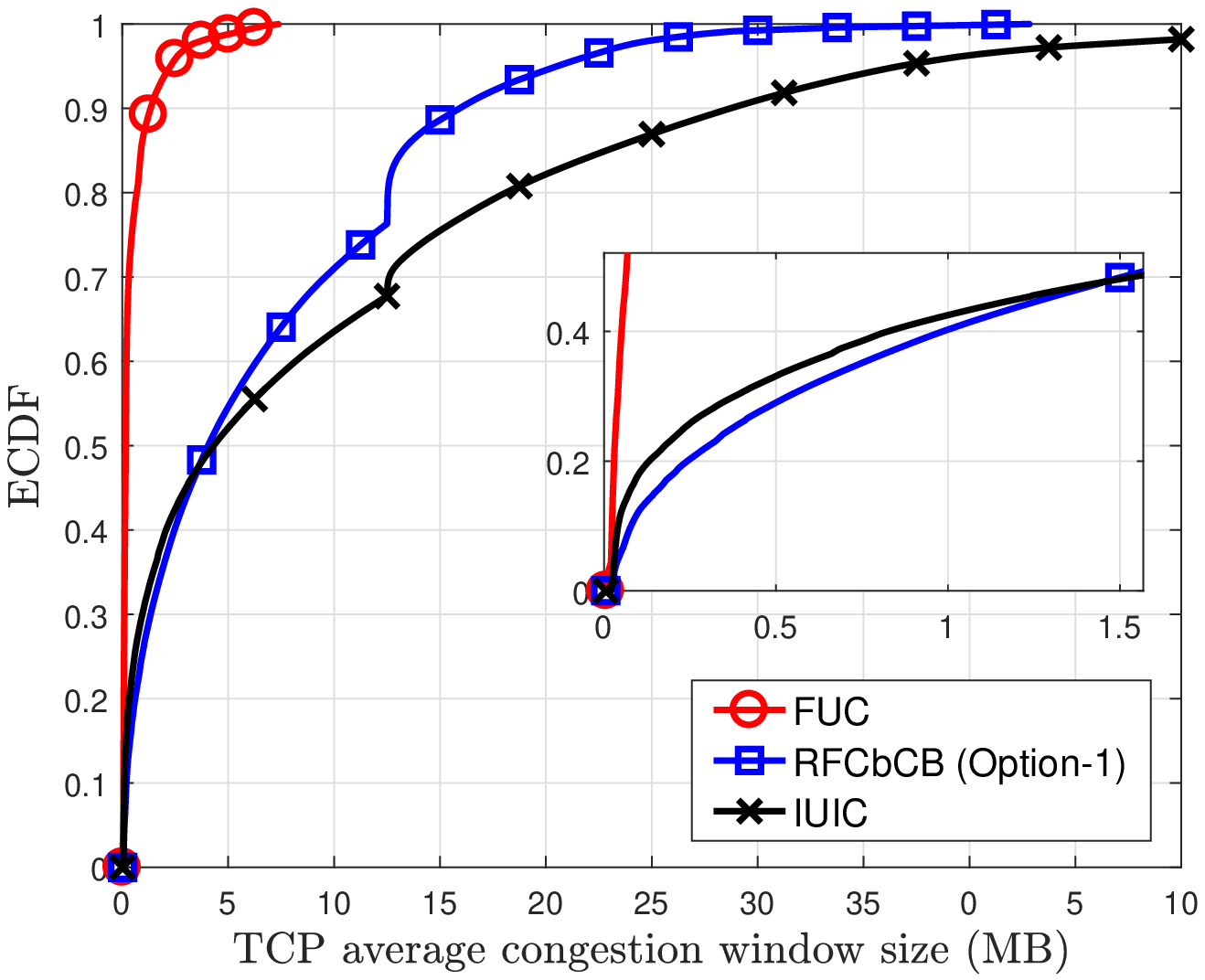}
\par\end{centering}
\centering{}{\small{}Fig. 5. TCP average congestion control performance.}{\small \par}
\end{figure}

Furthermore, the proposed RFCbCB scheme is demonstrated as best effort
since the minimum sub-frame misalignment threshold may not be satisfied
over all RFCs within the same requested $d_{c}:u_{c}$ sub-CB. Thus,
we investigate cases where the master cell slides to the nearest $d_{c}:u_{c}$
sub-CB to the requested one, i.e., RFCbCB (option-2), sacrificing
part of the TDD RFC flexibility due to the $d_{c}:u_{c}$ change.
Fig. 6 introduces the post-detection UL SINR, after the IRC decoding
as in eq. (3), of the RFCbCB (option-1) (slide only within requested
sub-CB), RFCbCB (option-2) (if applicable, slide to nearest sub-CB),
FUC, and IUIC, with UDP on top for DL:UL = 2:1. The RFCbCB (option-1)
provides substantial improvements in the UL SINR compared to the FUC,
i.e., an average of $+7.9$ dB increase. Moreover, RFCbCB (option-2)
further improves the perceived UL SINR level by an average of $+2.9$
dBs than RFCbCB (option-1), and with a bounded average loss of $-2.7$
dB to the optimal IUIC. 

Interestingly, the proposed RFCbCB (option-2) further improves the
overall DL and UL capacity per cluster, as depicted in Fig. 7, closely
approaching the UL-optimal IUIC. The RFCbCB (option-2) further significantly
reduces the probability of the CLI occurrence than RFCbCB (option-1)
by sliding to other RFC sub-CBs, at the expense of slightly changing
the traffic service ratio $d_{c}:u_{c}$, requested by slave cells.
Thus, the instantaneous UE rates may inflict a capacity loss, though,
being limited due to the conservative $d_{c}:u_{c}$ change. However,
on the average traffic statistics, the UL and DL traffic gets scheduled
and successfully decoded faster due to limited CLI, and thus, an enhanced
decoding ability, leaving more time and resources for incoming traffic.
As a result, the total UL and DL capacity per cell is further improved. 

\begin{figure}
\begin{centering}
\includegraphics[scale=0.52]{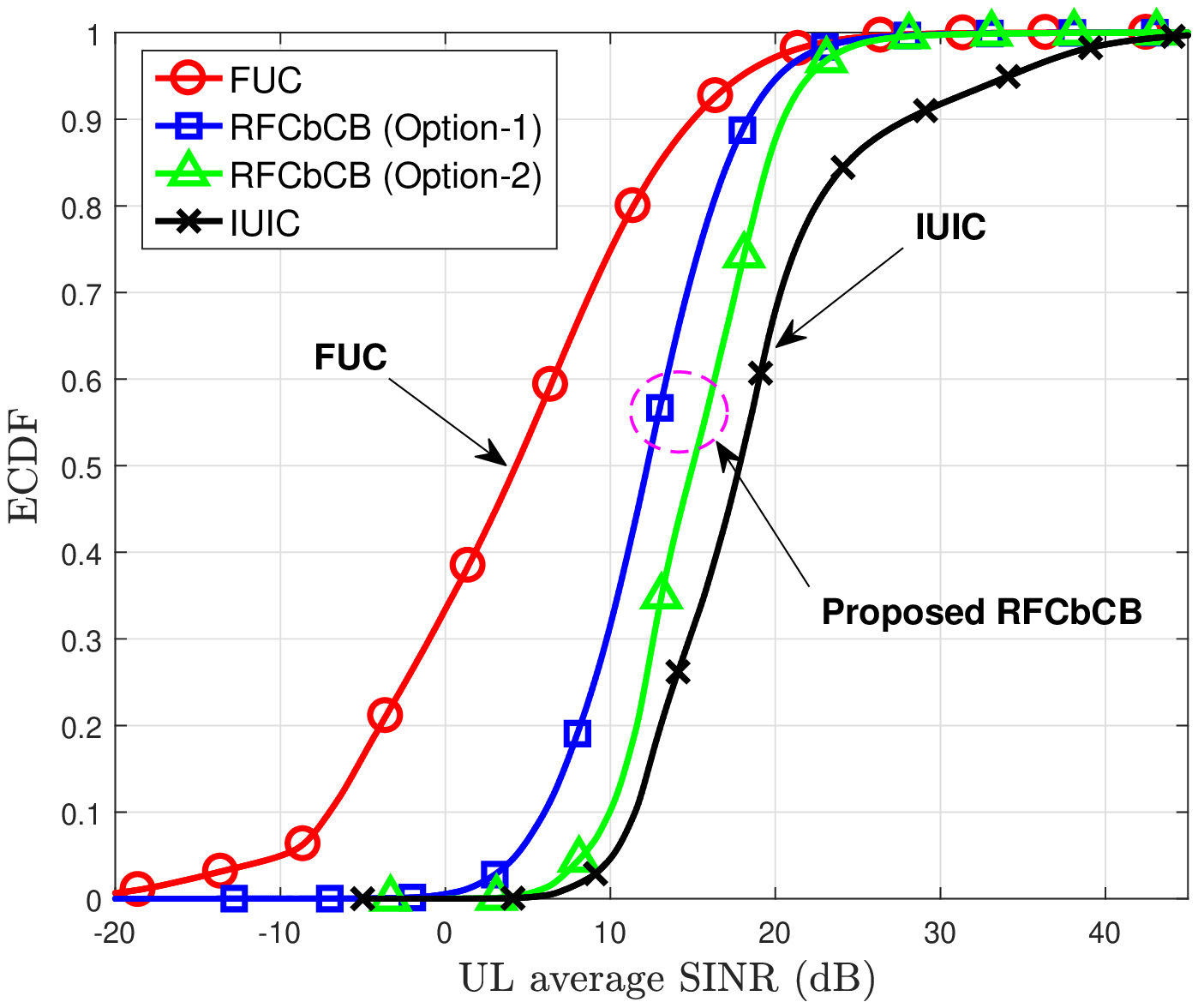}
\par\end{centering}
\centering{}{\small{}Fig. 6. UL average SINR performance (dB), with
UDP.}{\small \par}
\end{figure}
 
\begin{figure}
\begin{centering}
\includegraphics[scale=0.52]{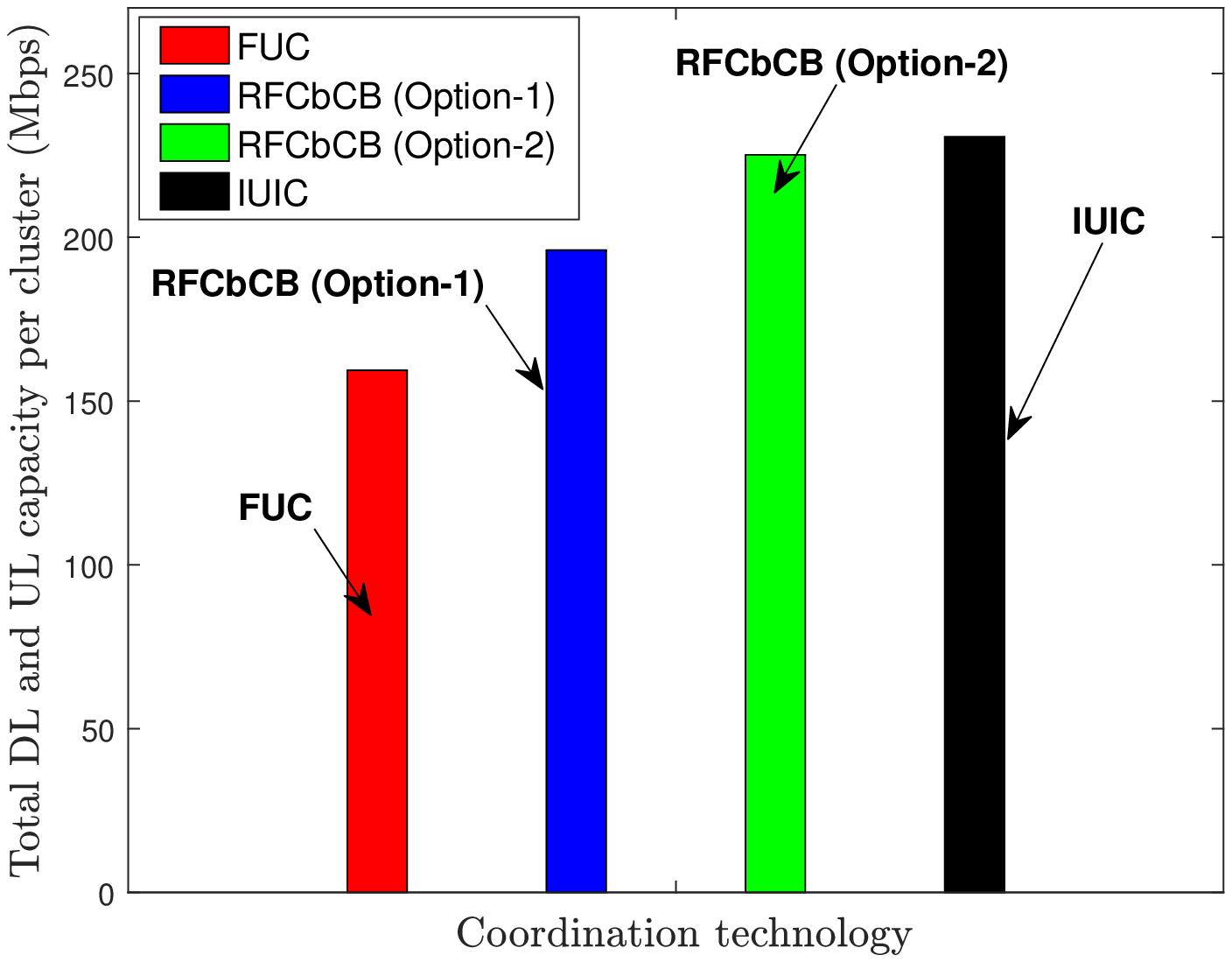}
\par\end{centering}
\centering{}{\small{}Fig. 7. Total DL and UL capacity per cluster
(Mbps), with UDP.}{\small \par}
\end{figure}

\section{Concluding Remarks }

In this work, a radio frame configuration based on sliding codebook
coordination scheme has been proposed for dynamic TDD 5G macro systems.
The proposed solution, with its introduced variations, offer a significant
capacity improvement, i.e., more than $\sim140.0\%$ gain, under both
TCP and UDP, and with a highly reduced inter-cell signaling overhead
size, limited to $\textnormal{B-bit}$. Compared to the state-of-the-art
TDD solutions from industry and academia, the proposed scheme has
been demonstrated as a flexible, high-performance and low-complexity
way to control the critical cross link interference (CLI) in dynamic
TDD networks. 

The main insights brought by this work are summarized as: (1) the
achievable capacity gains from the frame direction flexibility in
dynamic TDD macro systems can fully vanish or revert to a capacity
loss due to severe CLI, (2) the majority of the state-of-the-art dynamic
TDD coordination schemes assumes sophisticated inter-cell communications
to share the scheduling decisions, and transmission information. This
leads to a significant amount of control overhead, which is infeasible
in practice, and (3) proposed solution demonstrates a flexible coordination
scheme that dynamically exploits the fully dynamic TDD frame flexibility
when moderate levels of CLI are accepted. Otherwise, it slightly relaxes
the requirement of the fully flexible frame configuration, trading-off
an intended small capacity loss in the UE instantaneous rates for
the sake of a significant improvement in the overall capacity. A further
study with an analytical demonstration on the radio latency optimization
of the proposed solution will be conducted in a future work.

\section{Acknowledgments}

This work is partly funded by the Innovation Fund Denmark, Grant:
7038-00009B. Also, part of this work is performed in the framework
of the Horizon 2020 project ONE5G (ICT-760809) receiving funds from
the European Union.

\end{document}